\documentstyle[12pt]{article}
\newcommand{\beq}{\begin{equation}}
\newcommand{\eeq}{\end{equation}}
\newcommand{\la}[1]{\label{#1}}

\newcommand{\ci}[1]{\cite{#1}}
\newcommand{\ep}{\varepsilon}
\begin{document}
\title{Cluster reduction of the four-body Yakubovsky equations in 
configuration space for bound-state problem  and 
low-energy scattering } 
\author{S.L.\ Yakovlev${^{1,2,3}}$ and I.N.\ Filikhin${^{1}}$ }
\date{ }
\maketitle
\centerline{\small 
${^1}$ Department of Computational Physics, St. Petersburg State University,} 
\centerline{\small 198904 St. Petersburg Petrodvoretz Ulyanovskaya Str. 1, 
Russia}  
\centerline{\small 
${^2}$ Institute for Theoretical Physics, University of Groningen,} 
\centerline{\small
Nijenborgh 4, 9747 AG \ Groningen, The Netherlands }
\centerline{\small ${^3}$ E-mail: yakovlev@snoopy.niif.spb.su}
%
\abstract{\noindent 
A method using an expansion of the four-body Yakubovsky wave function
components onto the basis of the Faddeev-equation solutions for the 
two-cluster sub-Hamiltonian eigenfunctions is proposed. This expansion 
reduces the Yakubovsky differential equations to a system of coupled-channel 
equations for functions depending on the relative coordinates 
between the subsystems of the two-cluster partitions. On the basis of the 
resulting equations the four-nucleon bound-state problem and the 
zero-energy $n$-${^3}$H scattering problem are solved on the relatively
small computer.
}
\section{Introduction}

The Faddeev-Yakubovsky approach provides the correct way for a treatment
the few-body problem \ci{fad,yak}. It is based on the decomposition of
the wave function into components. This decomposition leads in the 
momentum representation to integral equations. The solutions of these 
equations are uniquely defined by inhomogeneous terms in contrast to 
the Lippmann-Schwinger equations for which this holds only in the 
two-body case. In the coordinate representation, the decomposition 
generates differential equations \ci{meryak} which are easier 
to solve numerically because they involve only interparticle potentials
and have simple boundary conditions \ci{mgl}-\ci{schelkok4}.

For the three-body case these equations are widely used for practical 
calculations. Unfortunately, already for the four-body systems, 
the numerical 
solution is very difficult and requires the most powerful supercomputer 
facilities \ci{cerba,glock}.

Recently, two new methods were proposed to solve the differential 
Faddeev-Yakubovsky equations (DYE) in configuration space
\ci{yakfil3,schelkok3,schelkok4,yakfil4}.
Both of them reduce the computational difficulties by several 
orders of  magnitude, so that the four-body equations can be solved
numerically on relatively small computers. One of the methods 
\ci{schelkok4} elaborated for the bound-state problem consists in the
systematic exploitation of the tensor structure of the matrices appearing
after discretization of  DYE with spline expansion and orthogonal 
collocation methods. This technique looks very appropriate to 
calculations of the four-body bound states producing more than five 
significant digits for the four-nucleon binding energy.

The key ingredient of the other method \ci{yakfil4} is the partial
diagonalization  of four-body DYE by making use of the basis 
of the eigenfunctions 
of Faddeev equations corresponding to subsystems of 
two-cluster fragmentations of the four-body system. Such expansions 
referred to conventionally as the coupled reaction channel (CRC) 
decomposition \ci{crcref} reduce DYE to a system of coupled 
equations for the functions depending only on the relative coordinates 
between the subsystems of two-cluster partitions. For this reason we name 
this procedure the cluster reduction.

The purpose of this paper is to give the systematic description of the 
latter method and to demonstrate its efficiency for numerical solutions
of both the four-nucleon bound-state and scattering-state problems.
The bound-state problem of the system $nnpp$ (${^4}$He) and the 
zero-energy $n$-${^3}$H scattering  problem have been chosen for the 
actual calculations. The MT I-III and S3
potential models have been used as 
the $N-N$ interactions \ci{schelkok4,glock,maltjo,afnan} . 
We would like to point out already
here that all the calculations were performed on a personal computer
with Intel DX-50 processor and 16 Mb of RAM. 

The paper is organized as follows. In the second section we recall those
portions of the papers \ci{meryak,meryakgig} which are required
to formulate the four-body bound-state and scattering-state problems 
on the base of DYE. In Sec. 3 the procedure of the partial 
diagonalization of DYE is described. There we give the appropriate  
definition of basis functions and discuss their main properties of 
completeness and biorthogonality. The resulting CRC equations
for the functions depending on the distance between the subsystems of the 
two-cluster partitions are used in the Sec. 4 for numerical
calculations of the four-nucleon bound-state (${^4}$He) and zero-energy
$n$-${^3}$H scattering problems. The last section summarizes the paper.
We have collected some technical details in the Appendix. 

\section{DYE formalism }

The four-particle wave function $\Psi$ should be decomposed into components
in one to one correspondence to all chains of partitions. The chains 
consist of two-cluster partitions $a_2$ ({\it e.g.}, $(ijk)l$ 
or $(ij)(kl)$) and three-cluster partitions 
$a_3$ ({\it e.g.}, $(ij)kl$) obeying the relation 
${a_3}\subset {a_2}$. The latter means that the partition $a_3$ can be 
obtained from the partition $a_2$ by splitting up one subsystem. It is 
easy to see that there exist $18$ chains of partitions for the 
four-particle system. 

The Yakubovsky wave-function components can be defined by the formulas  
\ci{yak,meryak,meryakgig}
\beq
\Psi_{a_{3}a_{2}} = R_{a_{3}}(E)V_{a_{3}}\sum_{(b_{3}\neq
a_{3})\subset a_{2}}R_{0}(E)V_{b_{3}}\Psi ,
\la{defcomp}
\eeq
where
$$
R_{a_{3}}(E) = (H_{0}+V_{a_{3}}-E)^{-1},\ R_{0}(E) = (H_{0}-E)^{-1}.
$$
Here $H_0$ is the kinetic-energy operator in the c.m. frame and 
$V_{a_3}$ stands for the two-particle potential acting inside the 
two particle subsystem of a partition $a_3$ ({\it e.g.}, $V_{a_3}$ = $V_{ij}$ 
with $a_3$ = $(ij)kl$).

If the function $\Psi$ is the solution of the Schr{\"o}dinger equation 
$$
(H_{0}+\sum_{a_3}V_{a_3}-E)\Psi = 0
$$ 
then the components $\Psi_{{a_3}{a_2}}$ obey the Yakubovsky equations
\cite{meryak,meryakgig}
$$
(H_{0}+V_{a_3}-E)\Psi_{{a_3}{a_2}}+V_{a_3}\sum_{({c_3}\neq {a_3})
\subset {a_2}}\Psi_{{c_3}{a_2}} = 
$$  
\beq
-V_{a_3}\sum_{{d_2}\neq {a_2}}\sum_{({d_3}\neq {a_3})\subset  
{a_2}}\Psi_{d_{3}d_{2}} . 
\la{yde}
\eeq 
The remarkable sum rule \ci{yak,meryak}
\beq
\Psi = \sum_{a_2}\sum_{a_3}\Psi_{a_{3}a_{2}} 
\la{totwf}
\eeq
allows one to construct the Schr{\"o}dinger equation solution $\Psi$. 

DYE in configuration space become  partial differential equations 
and require asymptotic boundary conditions to define a unique solution.
Let us now introduce the relative coordinates for the four-body problem
in order to fix the representation of operators in Eqs. (\ref{yde}) and 
to describe the asymptotics of the components $\Psi_{{a_3}{a_2}}$. 
There exist $18$ sets of relative Jacobi coordinates in one to one 
correspondence to $18$ chains of partitions ${a_3}{a_2}$.
Note, that among them the only two sets are topologically 
different. One corresponds to chains with partitions $a_2$ of the 
type $3+1$ ({\it e.g.}, (123)4) and the other one to chains with partitions
$a_2$ of the type $2+2$ ({\it e.g.}, (12)(34)). From now on we will assume that
all the particles have the same mass, say $m$. The relative coordinates
can be defined trough the radius vectors of particles by formulas 
 \ci{yakfil4,merfad}
\beq
\begin{array}{l}
{\bf x}_{a_3} = {\bf r}_{i} - {\bf r}_{j} \nonumber \\
{\bf y}_{{a_3},{a_2}} = ({\bf r}_{i}+{\bf r}_{j})/2 - {\bf r}_{k} \\
{\bf z}_{a_2} = ({\bf r}_{i}+{\bf r}_{j}+{\bf r}_{k})/3 - {\bf r}_{l} 
\nonumber
\end{array}
\la{coor3+1}
\eeq
for  $a_{2}=(ijk)l$, $a_{3}=(ij)kl$ and
\beq
\begin{array}{l}
{\bf x}_{b_3} = {\bf r}_{i} - {\bf r}_{j} \nonumber \\
{\bf y}_{{b_3},{b_2}} = {\bf r}_{k} - {\bf r}_{l} \\
{\bf z}_{b_2} = ({\bf r}_{i}+{\bf r}_{j})/2 - ({\bf r}_{k}+{\bf r}_{l})/2 
\nonumber
\end{array}
\la{coor2+2}
\eeq
for  $b_{2}=(ij)(kl)$, $b_{3}=(ij)kl$. With coordinates (\ref{coor3+1}) and 
(\ref{coor2+2}) the operators $H_0$ and $V_{a_3}$ have the form
$$
H_0 =
-\frac{\hbar^2}{m}(\Delta_{{\bf x}_{a_3}}+\alpha_{a_2}
\Delta_{{\bf y}_{{a_3},{a_2}}} 
+\beta_{a_2}\Delta_{{\bf z}_{a_2}}),
$$
$$
V_{a_3}=V_{a_3}({\bf x}_{a_3}),
$$
where $\Delta $ stands for Laplacian and 
$\alpha_{a_2}=3/4$, $\beta_{a_2}=2/3$ for the case of partitions 
$a_2$ of type $3+1$ and $\alpha_{a_2}=1$, $\beta_{a_2}=1/2$ for 
partitions $a_2$ of type $2+2$. 

The asymptotic boundary conditions for solutions of DYE (\ref{yde}) 
depend on the problem under consideration.  
For the bound-state problem one should look for
quadratically integrable solutions of Eqs. (\ref{yde}) satisfying 
\beq
\int{d{\bf x}_{a_3}d{\bf y}_{{a_3},{a_2}}d{\bf z}_{a_2}
|\Psi_{{a_3}{a_2}}({\bf x}_{a_3},{\bf y}_{{a_3},{a_2}},{\bf z}_{a_2})|^2} 
< \infty ,
\la{psiquadr}
\eeq
which will give a quadratically integrable wave-function 
$\Psi $,  
according to Eq.~ (\ref{totwf}). In fact, for the short-range potentials
the components $\Psi_{{a_3}{a_2}}$ have exponentially decreasing 
asymptotic behavior \ci{merfad}
\beq
\Psi_{{a_3}{a_2}}({\bf X}) 
\sim A_{{a_3}{a_2}}(\hat{X})\frac{
\exp\{ -\sqrt{|E|}|{\bf X}|\} }{|{\bf X}|^{5/2}} \ \  (|{\bf X}|
\rightarrow \infty ).     
\la{boundasy}
\eeq
In this equation we have introduced the shorthand notation ${\bf X}$
for the 9-vector 
$\{ {\bf x}_{a_3},{\bf y}_{{a_3,{a_2}}},{\bf z}_{a_2}\} $ and 
$\hat{X}$ for the unit vector ${\bf X}/|{\bf X}|$. For convenience
of notation, let us define six-vectors 
${\bf x}_{a_2} = \{ {\bf x}_{a_3}, {\bf y}_{{a_3},{a_2}}\} $ and 
${\bf y}_{a_3} = \{{\bf y}_{{a_3},{a_2}},{\bf z}_{a_2}\} $.
 The vectors ${\bf x}_{a_2}$ 
and
${\bf y}_{a_3}$ have a simple meaning. The first one describes
the relative positions of particles belonging to subsystems of the
partition $a_2$, and the second one corresponds to the configuration of 
particles $k,l$ and the center of mass of  the pair $(ij)$ for 
$a_3$ = $(ij)kl$. Note, that for the four-body position vector ${\bf X}$ 
we have the representations 
$$
{\bf X}= \{ {\bf x}_{a_3}, {\bf y}_{a_3}\} =\{ {\bf x}_{a_2}, {\bf z}_{a_2}\} 
. 
$$

In the case of the scattering problem we will deal with the functions marked 
by sign $+$. For these functions the asymptotic form of the components 
$\Psi_{{a_3}{a_2}}^{(+)}$  depends on the initial state of the system.  
We restrict ourselves to the scattering with two clusters in the initial 
state. Two-cluster channels can be characterized by the bound states of 
the two-cluster Hamiltonians
\beq 
H_{a_2} = T_{a_2}+\sum_{{a_3}\subset {a_2}}V_{a_3}\equiv 
-\frac{\hbar^2}{m}(\Delta_{{\bf x}_{a_3}}+\alpha_{a_2}
\Delta_{{\bf y}_{{a_3},{a_2}}}) + \sum_{{a_3}\subset {a_2}}V_{a_3}(
 {\bf x}_{a_3})
\la{ha2}
\eeq
describing the three-body subsystem $(ijk)$ for ${a_2}=(ijk)l$ and
the two noninteracting pairs $(ij)$ and $(kl)$ for ${a_2}=(ij)(kl)$.
We denote by $\phi^{a_3}_{a_2}({\bf x}_{a_2})$ the Faddeev components 
of the bound state of the Hamiltonian $H_{a_2}$ obeying the Faddeev 
equations
$$
( T_{a_2}+V_{a_3}-\ep_{a_2}) \phi^{a_3}_{a_2} = 
-V_{a_3}\sum_{({c_3}\neq {a_3})\subset {a_2}}\phi^{c_3}_{a_2}, 
$$
where $\ep_{a_2}$ is the respective binding energy.  

In order to characterize the three-cluster breakup channels let us 
introduce the two-body bound-state
wave function $\psi_{a_3}({\bf x}_{a_3})$ being the solution of the 
Schr{\"o}dinger equation
$$
( H_{a_3}-\ep_{a_3} ) \psi_{a_3}\equiv 
\{ -\frac{\hbar^2}{m}\Delta_{{\bf x}_{a_3}}+V_{a_3}({\bf x}_{a_3})-
\ep_{a_3}\} \psi_{a_3}({\bf x}_{a_3})=0.
$$
For the sake of simplicity of notation in what follows we will 
assume that the Hamiltonians $H_{a_2}$ and $H_{a_3}$ possess 
at most one bound state.

The four-body wave function components $\Psi^{(+)}_{{a_3}{a_2}}({\bf X},
{\bf p}_{l_2})$ for scattering with two clusters in the initial state 
can be arranged as the sum \ci{meryak, meryakgig}
\beq
\begin{array}{l}
\Psi^{(+)}_{{a_3}{a_2}}({\bf X},{\bf p}_{l_2}) = 
\delta_{{a_2}{l_2}}\phi^{a_3}_{
a_2}({\bf x}_{a_2})\exp \{ i{\bf p}_{a_2}\cdot {\bf z}_{a_2}\} \nonumber \\
+\phi^{a_3}_{a_2}({\bf x}_{a_2})U_{{a_2}{l_2}}({\bf z}_{a_2},{\bf p}_{l_2})+
\psi_{a_3}({\bf x}_{a_3})U^{a_2}_{{a_3}{l_2}}({\bf y}_{a_3},{\bf p}_{l_2})+
U_{{a_2}{a_3,{l_2}}}({\bf X},{\bf p}_{l_2}).
\end{array}
\la{scatasy}
\eeq
The first term in (\ref{scatasy}) is due to the initial state with two clusters 
being subsystems of the partition $l_2$ 
and moving with the relative momentum ${\bf p}_{l_2}$. The total 
energy $E$ and the momentum ${\bf p}_{l_2}$ are related by the expression 
$$
E=\ep_{l_2}+\beta_{l_2}\frac{\hbar^2}{m}{({\bf p}_{l_2})^{2}}.
$$
The second term in (\ref{scatasy}) describes the elastic (${a_2}={l_2}$)
and the rearrangement (${a_2}\neq {l_2}$) processes. 
For large separations, 
the functions  
$U_{{a_2}{l_2}}$ become spherical waves:   
\beq
U_{{a_2}{l_2}}({\bf z}_{a_2},{\bf p}_{l_2}) \sim {\cal A}_{{a_2}{l_2}}
\frac{\exp \{ i\sqrt{E-\ep_{a_2}}|{\bf z}_{a_2}|\} }{|{\bf z}_{a_2}|}
\ \ (|{\bf z}_{a_2}|\rightarrow \infty ),
 \la{u22}
\eeq
with ${\cal A}_{{a_2}{l_2}}$ being the amplitudes of the processes 
$(2\rightarrow 2)$. 
The remaining terms of (\ref{scatasy}) correspond to breakup processes 
$(2\rightarrow 3)$ and $(2\rightarrow 4)$. The asymptotics of the
functions $U$ have the form
\beq
U^{a_2}_{{a_3}{l_2}}({\bf y}_{a_3},{\bf p}_{l_2}) \sim {\cal A}^{a_2}
_{{a_3}{l_2}}
\frac{\exp \{ i\sqrt{E-\ep_{a_3}}|{\bf y}_{a_3}|\} }{|{\bf y}_{a_3}|^{5/2}}
\ \ (|{\bf y}_{a_3}|\rightarrow \infty ),
 \la{u23}
\eeq
\beq
U_{{a_2}{a_3},{l_2}}({\bf X},{\bf p}_{l_2}) \sim {\cal A}_{{a_2}{a_3},{l_2}}
\frac{\exp \{ i\sqrt{E}|{\bf X}|\} }{|{\bf X}|^{4}}
\ \ (|{\bf X}|\rightarrow \infty ).
 \la{u24}
\eeq
Here, the amplitudes ${\cal A}$ are related to the breakup amplitudes by 
the formulas
$$
f_{{a_3}{l_2}}(2\rightarrow 3) = \sum_{{a_2}\supset {a_3}}
{\cal A}^{a_2}_{{a_3}{l_2}},
$$
$$
f_{l_2}(2\rightarrow 4) = \sum_{a_2}\sum_{a_3}
{\cal A}_{{a_2}{a_3},{l_2}}.
$$

The asymptotic boundary conditions (\ref{scatasy}-\ref{u24}) make the 
solution of DYE (\ref{yde}) unique. For the more complicated case of 
scattering states with three and four clusters in the initial state 
we refer to the paper \ci{yakov}.

\section{Cluster reduction of the DYE}

In this section we consider the four-body bound state problem and the 
low-energy scattering problem when only the two-cluster channels are open. 
The latter
means that the energy of the system obeys the inequalities
$$
E-\ep_{a_2}\geq 0,\ E-\ep_{a_3} < 0,\  E < 0.
$$

Under these conditions the two last terms in (\ref{scatasy}) describing
virtual breakup processes vanish exponentially as $|{\bf X}|$ approaches  
$\infty$ and the asymptotics of $\Psi^{(+)}_{{a_3}{a_2}}$
becomes simpler. Note, that every component 
$\Psi_{{a_3}{a_2}}$ for both bound and scattering state now is an 
exponentially  decreasing function of the coordinates ${\bf x}_{a_2}$ 
corresponding to the separation of the particles inside the subsystems 
of the partition $a_2$. It means, that the essential part of the component
$\Psi_{{a_3}{a_2}}$ is concentrated in a tube-like domain of
the configuration space such that $|{\bf x}_{a_2}|\leq R_{a_2}$ with 
sufficiently large 
parameter $R_{a_2}$. It is natural to find the 
approximate 
solution of Eqs. (\ref{yde}) by imposing the following boundary
conditions: 
$$
\Psi_{{a_3}{a_2}}|_{\Gamma_{a_2}} = 0,
$$
where $\Gamma_{a_2}$ means the boundary of the cylinder: 
$|{\bf x}_{a_2}| = R_{a_2}$. 
Imposing these boundary conditions is the 
first step of the reduction. As a second step, we expand the components 
in the following series:  
\beq
\Psi_{{a_3}{a_2}}({\bf x}_{a_2},{\bf z}_{a_2}) = 
\sum_{l=0}^{\infty}\phi_{{a_2},l}^{a_3}({\bf x}_{a_2})
F_{a_2}^{l}({\bf z}_{a_2}).
\la{compdecomp}
\eeq
Here $F^{l}_{a_2}({\bf z}_{a_2})$ are new, unknown, functions and 
$\phi^{a_3}_{{a_2},l}({\bf x}_{a_2})$ are the eigenfunctions of the Faddeev
equations for the subsystems of the partition $a_2$ 
\beq
(T_{a_2} + V_{a_{3}})\phi_{{a_2},l}^{a_3} + V_{a_3}
\sum_{({c_3}\neq {a_3})\subset {a_2}}\phi_{{a_2},l}^{c_3} = 
\ep_{a_2}^{l}\phi_{{a_2},l}^{a_3},
\la{fadbaseq}
\eeq
obeying the boundary conditions
\beq
\phi_{{a_2},l}^{a_3}({\bf x}_{a_2})|_{\Gamma_{a_2}} = 0.
\la{fadbasbound}
\eeq
The equations (\ref{fadbaseq}) with the spatially restricted conditions 
(\ref{fadbasbound}) are known to have a purely discrete spectrum of
real eigenvalues $\ep^{l}_{a_2}$ \ci{ewhof}. The functions
$\phi_{{a_2},l}^{a_3}$ and the eigenvalues $\ep^{l}_{a_2}$ are assumed to be 
enumerated in order of growth the quantities $\ep^{l}_{a_2}$. One observes, 
that for large enough value of $R_{a_2}$, the functions 
$\phi_{{a_2},0}^{a_3}$ and the quantity $\ep^{0}_{a_2}$ approach the 
ground-state Faddeev components and the bound-state  
energy of the Hamiltonian 
(\ref{ha2}), respectively. The remaining part of basis functions extends 
the function 
$\phi_{{a_2},0}^{a_3}$ up to a complete set \ci{ewhof}. This set of 
eigenfunctions is not orthogonal due to the nonsymmetry of 
(\ref{fadbaseq}) with respect to the superscript $c_3$. The biorthogonal
basis is formed \ci{ewhof} by the solutions of the adjoint to Eqs. 
(\ref{fadbaseq}): 
\beq
(T_{{a_2}} + V_{a_{3}})\psi_{{a_2},k}^{a_3} + 
\sum_{({c_3}\neq {a_3})\subset {a_2}}V_{c_3}\psi_{{a_2},k}^{c_3} = 
\ep_{a_2}^{k}\psi_{{a_2},k}^{a_3}.
\la{adfadbaseq}
\eeq
The completeness and biorthogonality properties for the functions 
$\phi_{{a_2},l}^{a_3}$ and $\psi_{{a_2},k}^{a_3}$ read
$$
\sum_{k=0}^{\infty }
\phi_{{a_2},k}^{a_3}({\bf x}_{a_2})\psi_{{a_2},k}^{c_3}({\bf x}_{a_2}
^{\prime })={\delta_{{a_3}{c_3}}}\delta ({\bf x}_{a_2}-{\bf x}_{a_2}
^{\prime }),  
$$
$$
\sum_{{a_3}\subset {a_2}}\int d{{\bf x}_{a_2}}\phi_{{a_2},l}^{a_3}
({\bf x}_{a_2})\psi_{{a_2},k}^{c_3}({\bf x}_{a_2})= \delta_{lk}. 
$$
Using these conditions, we express the coefficients in
(\ref{compdecomp}) as the projections
\beq 
F^{l}_{a_2}({\bf z}_{a_2})=\sum_{{a_3}\subset {a_2}}\int 
d{{\bf x}_{a_2}}\psi_{{a_2},l}^{a_3}({\bf x}_{a_2})\Psi_{{a_3}{a_2}}
({{\bf x}_{a_2}},{{\bf z}_{a_2}}).
\la{projcomp}
\eeq

Substituting the expansions (\ref{compdecomp}) into the DYE (\ref{yde})
and projecting onto the biorthogonal basis functions 
$\psi_{{a_2},l}^{a_3}({\bf x}_{a_2})$ we get the final equations for the 
coefficients $F^{k}_{a_2}({\bf z}_{a_2})$:  
$$
(-\frac{\hbar^{2}}{m}{\beta_{a_2}}\Delta_{{\bf z}_{a_2}} - E + 
\ep_{a_2}^{k})F_{a_2}^{k}(
{\bf z}_{a_2}) =   
$$
\beq
-\sum_{{a_3}\subset {a_2}}\langle 
\psi^{a_3}_{a_2,k}|V_{a_3}\sum_{{d_2}\neq {a_2}}\sum_{({d_3}\neq
{a_3})\subset {a_2}}\sum_{l\geq
0}\phi_{{d_2},l}^{{d_3}}({\bf x}_{d_2})F_{d_2}^{l}({\bf z}_{d_2})\rangle 
\la{redyakeq}.
\eeq
Here the brackets $\langle \cdot |\cdot \rangle $
stand for the integration over ${\bf x}_{a_2}$ in the domain 
$|{\bf x}_{a_2}|\leq ~R_{a_2}$ and the coordinates ${\bf x}_{d_2}$ and 
${\bf z}_{d_2}$ are assumed to be expressed through the set 
${\bf x}_{a_3}$, ${\bf y}_{{a_3},{a_2}}$, ${\bf z}_{a_2}$ 
\ci{merfad,yakov}.

The asymptotic boundary conditions for the functions 
$F^{k}_{a_2}({\bf z}_{a_2})$ can be easily obtained from
(\ref{psiquadr}) and (\ref{scatasy}-\ref{u24}) by projecting  
according to formula (\ref{projcomp}). For the bound state-problem
they read
\beq
\sum_{k \geq 0}
\int d{{\bf z}_{a_2}}|F^{k}_{a_2}({\bf z}_{a_2})|^{2} < \infty ,
\la{redbound}
\eeq
and for the scattering states they have the following  `two-body'
form
\beq
F_{a_2}^{0}({\bf z}_{a_2}) \sim \delta_{{a_2}{l_2}}\exp
\{ i{\bf p}_{a_2}\cdot {\bf z}_{a_2}\}  
+{\cal A}_{{a_2}{l_2}}|{\bf z}_{a_2}|^{-1}\exp\{ i\sqrt{E-\ep^{0}_{a_2}}
|{\bf z}_{a_2}|\} 
\la{redscatopen}
\eeq
for the open channels and
\beq
F_{a_2}^{k}({\bf z}_{a_2}) \sim 0, \ k\geq 1
\la{redscatclosed}
\eeq
for the closed channels.
The equations (\ref{redyakeq}) are the desired coupled channel equations
for four distinguishable particles.

Let us now turn to the case of identical particles. On the first stage,
note that due to the identity of particles, it suffices to define  
only two Yakubovsky components $\Psi_{{a_3}{a_2}}$, namely, one for the 
partition $a_2$ of the type $3+1$ and one for the partition $a_2$ of 
the type $2+2$. We fix these two components choosing $a_3 = (12)34$, 
$a_2 = (123)4$ and $b_3 = (12)34$, $b_2 = (12)(34)$ and denoting 
$\Psi_{{a_3}{a_2}}= \Psi_1$, $\Psi_{{b_3}{b_2}}=\Psi_2$. 
 
Let $P^{\pm}$ be the cyclic and anticyclic permutation operators for 
four particles and $P^{\pm}_i$ be the same for a three-particle subsystem 
where the subscipt $i$ corresponds to the fourth particle, that is 
not participating in the permutation. With these operators the 
representation 
(\ref{totwf}) can be rewritten in terms of the components $\Psi_1$ and 
$\Psi_2$ as follows
$$
\Psi = [I+\sigma {P^+} + {P^+}{P^+} + \sigma {P^-}]
[I+{P^{+}_{4}} + {P^{-}_{4}}]\Psi_1 +
[I+{P^{+}_{1}} + {P^{-}_{1}}][I + {P^+}{P^+}]\Psi_2.
$$
The DYE (\ref{yde}) for the functions $\Psi_{i}$, $i=1,2$ read
\begin{eqnarray} 
(H_0 + V - E)\Psi_1 + V(P^{+}_{4} + P^{-}_{4})\Psi_1 &=&
-V[(P^{+}_{1} + \sigma P^{+})
\Psi_{1} + (P^{+}_{1} + P^{+}_{4})\Psi_{2} ],\nonumber \\ 
(H_0 + V - E)\Psi_2 + V(P^{+}P^{+})\Psi_2 &=&  
-V[P^{+} + \sigma P^{+}_{1}]P^{+}\Psi_{1}.  \nonumber 
\end{eqnarray} 
Here $V\equiv V({\bf x}_{12})$ and $\sigma = +1(-1)$ for the boson (fermion) 
case.

The expansion (\ref{compdecomp}) assumes the form 
\beq
\Psi_{i} ({\bf x}_{i}, {\bf y}_{i} ) = \sum_{l=0}^{\infty } 
\phi^{i}_{l}({\bf x}_{i})F^{l}_{i}({\bf z}_{i}), \ \ i=1,2, 
\la{symexp}
\eeq
with the basis functions $\phi^{i}_{l}$ being the solutions of 
the symmetrized Faddeev equations 
$$
(T_1 + V)\phi^{1}_{l} + V(P^{+}_{4} + P^{-}_{4})\phi^{1}_{l} = 
\ep^{l}_{1}\phi^{1}_{l},
$$
$$
(T_2 + V)\phi^{2}_{l} + V P^{+}P^{+}\phi^{2}_{l} = \ep^{l}_{2}\phi^{2}_{l}.
$$
The biorthogonal basis functions $\psi^{k}_{l}$ are the solutions of 
the respective adjoint equations 
$$
(T_1 + V)\psi^{1}_{l} + (P^{+}_{4} + P^{-}_{4})V\psi^{1}_{l} = 
\ep^{l}_{1}\psi^{1}_{l},
$$
$$
(T_2 + V)\psi^{2}_{l} +  P^{+}P^{+}V\psi^{2}_{l} = \ep^{l}_{2}\psi^{2}_{l}.
$$
The expansion (\ref{symexp}) leads to the symmetrized form of 
(\ref{redyakeq})
\newpage 
$$
\left\{ -\frac{\hbar^{2}}{m}\frac{2}{3}\Delta_{{\bf z}_{1}} - E
+\ep^{k}_{1} \right\} 
F^{k}_{1}({\bf z}_{1}) = -\langle \psi^{1}_{k}|V[P^{+}_{1} + \sigma P^{+}]
\sum_{l\geq 0}\phi^{1}_{l}F^{l}_{1}\rangle 
$$
\beq
- \langle \psi^{1}_{k}|V[P^{+}_{1} + P^{+}_{4}]
\sum_{m\geq 0}\phi^{2}_{m}F^{m}_{2}\rangle , 
\la{redyakeqsym}
\eeq 
$$
\left\{ -\frac{\hbar^{2}}{m}\Delta_{{\bf z}_{2}} - E +\ep^{k}_{2}\right\} 
F^{k}_{2}({\bf z}_{2}) =
-\langle \psi^{2}_{k}|V[P^{+}_{1} + \sigma P^{+}_{1}]P^{+}
\sum_{l\geq 0}\phi^{1}_{l}F^{l}_{1}\rangle .
$$
The boundary conditions for the equations (\ref{redyakeqsym}) follow 
directly from (\ref{redbound}), (\ref{redscatopen}), and 
(\ref{redscatclosed}).

\section{Application to the four-nucleon system}

To apply the equations (\ref{redyakeqsym}) to the four-nucleon system 
one needs to perform a partial-wave analysis. Taking into account the 
conservation of the total angular momentum $L$ and the total spin $S$ of the 
system for chosen $N-N$ forces S3 and MT I-III we can use the $L-S$
coupling scheme and obtain for the $s-$wave components
\beq
F^{k}_{S,l}(z) = |{\bf z}|\int d{\hat z}F^{l}_{k}({\bf z}), \ \ 
k=1,2 
\la{swaveF}
\eeq 
(here and in what follows the index $S$ is introduced to distinguish the 
states with the total spin $S=0$ or $S=1$) the following resulting 
$s-$wave equations\footnote{  
We have restricted the summation in formula (\ref{symexp})
to finite numbers $N_1$ and $N_2$, so that (\ref{swaveyak}) should be 
considered as approximate equations,which become exact in the limit 
${N_1},{N_2}\rightarrow \infty $. } 
$$
(-\frac{2}{3}\partial^{2}_{z} -\ep +\ep^{1}_{S,m})F^{1}_{S,m}(z) =  
$$
$$
-\frac{1}{2}\langle \Psi^{1}_{S,m}(x,y)|V^{S}_{1}(x)\int_{-1}^{1}du
\int_{-1}^{1}dv\frac{xyz}{{x_3}{y_3}{z_3}}D^{S}\sum_{l=0}^{N_1}
\Phi^{1}_{S,l}({x_3},{y_3},)F^{1}_{S,l}({z_3})\rangle 
$$
\beq
-\frac{1}{2}\langle \Psi^{1}_{S,m}(x,y)|V^{S}_{1}(x)\int_{-1}^{1}du
\int_{-1}^{1}dv\frac{xyz}{{x_4}{y_4}{z_4}}C^{S}_{1}\sum_{s=0}^{N_2}
\Phi^{2}_{S,s}({x_4},{y_4})F^{2}_{S,s}({z_4})\rangle ,
\la{swaveyak}
\eeq
$$
(-\frac{1}{2}\partial^{2}_{z}-\ep +\ep^{2}_{S,n})F^{2}_{S,n}(z) = 
$$
$$
-\langle \Psi^{2}_{S,n}(x,y)|V^{S}_{2}(x)
\int_{-1}^{1}dv\frac{xyz}{{x_5}{y_5}{z_5}}C^{S}_{2}\sum_{l=0}^{N_1}
\Phi^{1}_{S,l}({x_5},{y_5})F^{1}_{S,l}({z_5})\rangle .
$$
Here the brackets $\langle \cdot |\cdot \rangle $ mean the 
integration over $x$ and $y$ in the domain $x^{2}+y^{2}\leq R^{2}$ 
and summation over the components of vector-functions $\Phi^{k}_{S,l}$ 
and $\Psi^{k}_{S,m}$ being the solutions of $s-$wave Faddeev equations
\newpage 
$$
\left\{ -\partial^{2}_{x} -\frac{3}{4}\partial^{2}_{y} + V^{S}_{1}(x) 
\right\} 
\Phi^{1}_{S,l}(x,y) + V^{S}_{1}(x)\int_{-1}^{1} dv \frac{xy}{{x_1}{y_1}}
B^{S}_{1}\Phi^{1}_{S,l}({x_1},{y_1}) = 
$$
$$
=\ep^{1}_{S,l}\Phi^{1}_{S,l}(x,y),
$$
\beq
\left\{ -\partial^{2}_{x} - \partial^{2}_{y} + V^{S}_{2}(x) \right\} 
\Phi^{2}_{S,n}(x,y) + V^{S}_{2}(x)
B^{S}_{2}\Phi^{2}_{S,n}({x_2},{y_2}) = 
\la{swavefad}
\eeq
$$
=\ep^{2}_{S,n}\Phi^{2}_{S,n}(x,y),
$$
and adjoint equations
$$
\left\{ -\partial^{2}_{x} -\frac{3}{4}\partial^{2}_{y} + V^{S}_{1}(x) 
\right\} 
\Psi^{1}_{S,l}(x,y) + \int_{-1}^{1} dv \frac{xy}{{x_1}{y_1}}
B^{S}_{1}V^{S}_{1}({x_1})\Psi^{1}_{S,l}({x_1},{y_1}) = 
$$
$$
=\ep^{1}_{S,l}\Psi^{1}_{S,l}(x,y),
$$
\beq
\left\{ -\partial^{2}_{x} - \partial^{2}_{y} + V^{S}_{2}(x) \right\} 
\Psi^{2}_{S,n}(x,y) + 
B^{S}_{2}V^{S}_{2}({x_2})\Psi^{2}_{S,n}({x_2},{y_2}) = 
\la{swaveadj}
\eeq
$$
=\ep^{2}_{S,n}\Psi^{2}_{S,n}(x,y).
$$
In the above equations we have renormalized the energy of the system 
according to 
 $\ep = mE/{\hbar^2}$. The matrices $B^{S}_{k}$, $D^S$ and 
$C^{S}_{k}$, $k=1,2$ realize the representation of operators 
$P^{+}_{4}+P^{-}_{4}$, $P^{+}P^{+}$, $P^{+}_{1}-P^{+}$, $P^{+}_{1}+
P^{+}_{1}$, and $(P^{+}-P^{+}_{1})P^{+}$ in the spin-isospin space of the 
four-nucleon system, respectively. We give their values for the systems $nnpp$ and 
$nnnp$ in the Appendix. There we have described the representations 
of interaction potentials $V^{S}_{k}$ and values of coordinates 
${x_i},{y_i},{z_i}$ appearing in the equations.

The components $F^{k}_{S,l}(z)$, due to (\ref{swaveF}), obey the regularity
conditions 
$$
F^{k}_{S,l}(0)=0.
$$
The asymptotic boundary conditions for the bound-state problem 
according to (\ref{redbound}) and (\ref{swaveF}) have the form
\beq
\sum_{l\geq 0} \int_{0}^{\infty }d z |F^{k}_{S,l}(z)|^{2} < \infty , \ \ \ 
k=1,2.
\la{F2}
\eeq
For the $n$-${^3}$H scattering problem the only three-body subsystems 
have the bound state, so that (\ref{redscatopen}) and
(\ref{redscatclosed}) lead to the conditions
$$
F^{1}_{S,0}(z,p) \sim p^{-1}\sin{pz} + a_{S}(p)\cos{pz},   
$$
\beq
F^{1}_{S,l}(z,p) \sim 0, \  l\geq 1,  
\la{Fscat}
\eeq
$$
F^{2}_{S,l}(z,p) \sim 0, \  l\geq 0.   
$$
For the zero relative $n$-${^3}$H scattering energy the first condition
of  (\ref{Fscat}) should be replaced by
\beq
F^{1}_{S,0}(z,0) \sim z-A_{S},
\la{zeroscat}
\eeq
where $A_{S}$ is the $n$-${^3}$H scattering length. The relationship
between the scattering amplitude $a_{S}(p)$ and the phase shift
$\delta_{S}(p)$ is of the form 
\beq
a_{S}(p) = \frac{\tan{\delta_{S}(p)}}{p}. 
\la{phaseshiftdef}
\eeq
Alternatively, the scattering length $A_{S}$ can be obtained from
$a_{S}(p)$ by taking the limit,  
\beq
A_{S} = - \lim_{p\rightarrow 0} a_{S}(p).
\la{extrapol}
\eeq

The equations (\ref{swaveyak}) were solved numerically by making 
use of a finite-difference approximation for the differential operator
$\partial^{2}_{z}$ and a spline expansion for the integrands on the 
right hand-side. The mesh parameters were varied extensively to obtain
the relative numerical uncertainty less than $1\% $. The basis functions
$\Phi^{k}_{S,l}$ and $\Psi^{k}_{S,l}$ were constructed numerically as 
the solutions of the eigenvalue problems (\ref{swavefad}) and 
(\ref{swaveadj}). The maximal nonorthogonality of the basis functions caused
by the numerical solutions of  (\ref{swavefad}) and 
(\ref{swaveadj}) was given by 
$$
\langle \Psi^{k}_{S,l}|\Phi^{k}_{S,m} \rangle \leq 10^{-6}, \ l \neq m. 
$$

We have used 
S3 and MT I-III $s-$wave potentials for the bound-state calculations. 
For $n$-$^3$H scattering we have used only the MT I-III potential.
These potentials 
are known to have a strong infinite (MT I-III) and finite (S3) 
repulsive core. Such a singular behavior of the potentials requires an 
appropriate treatment of the problem, so we have chosen these potentials 
to demonstrate the efficiency of our approach. On the other hand, the 
data of the direct (except for numerical approximations) solutions 
of the four-nucleon Yakubovsky equations (YE) for bound states with these 
potentials are available \ci{schelkok4,glock} 
which facilitates 
the comparison of the results of calculations. 

In Table 1 we show the convergence of our calculations for ${^4}$He 
binding energy with respect to the number of terms taken into account 
in the expansion (\ref{symexp}) and  in Eqs. (\ref{swaveyak}),
respectively. 
As one can see a converged result within the uncertainty less than 
$1\% $ can be reached with about $10$ basis functions. In Table 2  
we have collected our final values for ${^4}$He binding energy and the data 
of solutions of YE  \ci{schelkok4,glock,tjon}.  The agreement of 
calculations is near perfect.

The calculations of $n$-${^3}$H scattering were performed for 
the zero neutron laboratory energy. We have used 
 the boundary conditions  
(\ref{zeroscat})  for zero energy calculations. The scattering 
 length $A_S$ was extracted from the asymptotics 
of the solutions according to the formula (\ref{zeroscat}).
 We have collected the results of our calculations
for spin-singlet and spin-triplet scattering lengths $A_0$ and
$A_1$ together with available data of the other authors and
experimental values in Table 3. Comparison 
of the results is hampered by the fact that  different approaches 
and different models for $N$ - $N$ forces were used. Nevertheless, 
the agreement 
of the results is reasonable except for the case of the paper \ci{heiss} 
where 
the Resonating Group Method (RGM) has been used. In our notations, the 
RGM ansatz corresponds to taking into account only one basis 
state in (\ref{symexp}). We have performed the calculations within this 
condition and obtained 3.41 fm for the spin-singlet scattering length 
$A_0$ which is very close to the result of \ci{heiss}. Hence, we can 
suppose   
that the difference between our converged result and the RGM one is due to 
the approximations of RGM.

\section{Summary}

We have described the cluster reduction method to treat the four-body 
bound-state and scattering-state problem. The method leads to 
CRC equations for the functions depending on the relative coordinates 
between the subsystems of the two-cluster partitions. After a suitable
partial-wave decomposition, these equations become the one-dimensional
integro-differential equations which can be solved numerically on 
rather small computers in contrast to the original DYE, for which numerical
solution even for simple $N-N$ forces requires supercomputer
facilities. The results of calculations given in the previous section
show the efficiency of  reduced DYE (\ref{redyakeqsym}) and 
(\ref{swaveyak})
for the numerical solution of both the bound-state and scattering-state
problems for the four-nucleon system. 
	The method proposed can be extended both to the case of more realistic
nucleon-nucleon potentials in four-nucleon system, and to systems
consisting of more than four particles. In the latter case the N-body 
differential Yakubovsky equations \ci{meryak}
should be taken as a starting point for the reduction procedure.

The efficiency of the cluster-reduction method for the scattering problem 
demonstrated for the $n$-${^3}$H system suggests it will be 
useful for the direct four-body calculations of various multichannel 
reactions.

\vskip1cm
\noindent {\Large {\bf Acknowledgments}}
\\ 

\noindent 
This work was finished during the stay one of the authors
(S.L.Y.) at the Institute for Theoretical Physics University of 
Groningen made possible under the bilateral agreement between 
Groningen University and University of Sankt Petersburg for scientific 
exchange. The authors are thankful to Prof. L.P. Kok for fruitful
discussions, careful reading of the manuscript and numerous suggestions 
for improving the paper. 
This work was supported in part by INTAS Grant 
INTAS-93-1815 and RFBR Grant No. 96-02-17021-a. 
\newpage  

\noindent {\Large {\bf Appendix}}

\noindent 
In this Appendix we describe the representation of the potentials 
$V^{S}_{k}$ and 
the matrices $B^{S}_{k}$, $D^S$, and $C^{S}_{k}$, $k=1,2$ and give
the formulas for the coordinates ${x_i},{y_i}, {z_i}$ appearing  
in the integrals of Eqs. (\ref{swaveyak}), (\ref{swavefad}), and 
(\ref{swaveadj}).
	For the bound state of the $nnpp$ system the total spin equals 
to zero, $S=0$. The operators $V^{0}_{k}$ are the diagonal matrices
such that
$$
V^{0}_{1}=V^{0}_{2} = diag\{ v^{s}, v^{t}\} ,
$$
where $v^{s}$ and $v^{t}$ stand for the singlet and triplet $N-N$ 
potentials, respectively.
The matrices $B^{0}_{k}$, $D^{0}$ and $C^{0}_{k}$ have the form
$$
B^{0}_{1} = D^{0} = C^{0}_{1} = \left( \begin{array}{rr}
\frac{1}{4} & -\frac{3}{4} \\ 
-\frac{3}{4} & \frac{1}{4} 
\end{array} \right) ,
$$
$$
B^{0}_{2} =  C^{0}_{2} = \left( \begin{array}{rr}
1  & 0   \\ 
0  & 1  
\end{array} \right) .
$$

For the system $n$-${^3}$H the values of the total spin $S$ equal 0 or 1.
The operators $V^{S}_{k}$ are the diagonal matrices 
$$
V^{0}_{1} = diag\{ v^{s}, v^{t}, v^{s} \} ,  
$$
$$
V^{0}_{2} = v^{s} , 
$$
$$
V^{1}_{1} = diag\{ v^{s}, v^{t}, v^{s}, v^{t} \} ,  
$$
$$
V^{1}_{2} = diag\{ v^{s}, v^{t} \} . 
$$
The explicit form of the matrices  $B^{S}_{k}$, $D^S$, and $C^{S}_{k}$ is
as follows 
$$
B^{0}_{1} = \left( \begin{array}{rrr} 
\frac{1}{4} & -\frac{3}{4} & 0 \\ 
-\frac{3}{4} & \frac{1}{4}  & 0 \\ 
0 & 0 & -\frac{1}{2}
\end{array} \right) , 
$$
$$
B^{0}_{2} = -1 
$$
$$
B^{1}_{1} = \left( \begin{array}{rrrr}
\frac{1}{4} & -\frac{3}{4} & 0 & 0 \\
-\frac{3}{4} & \frac{1}{4} & 0 & 0 \\
0 & 0 & -\frac{1}{2} & 0 \\
0 & 0 & 0 & -\frac{1}{2} 
\end{array} \right) ,
$$
$$
B^{1}_{2} = \left( \begin{array}{rr}
0 & 1 \\
1 & 0 
\end{array} \right) ,
$$
$$
D^{0} = \left( \begin{array}{rrr}
\frac{1}{12} & \frac{3}{4} & \frac{1}{6}{\sqrt 2} \\
-\frac{1}{4} & -\frac{1}{4} & -\frac{1}{2}{\sqrt 2} \\
\frac{1}{3}{\sqrt 2} & \frac{1}{6} & 0 
\end{array} \right) ,
$$
$$
D^{1} = \left( \begin{array}{rrrr}
-\frac{1}{12} & \frac{1}{4} & -\frac{1}{6}{\sqrt 2} & \frac{1}{2}{\sqrt 2} \\ 
\frac{1}{4} & -\frac{1}{12} & \frac{1}{2}{\sqrt 2}& -\frac{1}{6}{\sqrt 2} \\ 
\frac{1}{3}{\sqrt 2} & 0 & -\frac{1}{6} & 0 \\
0 & \frac{1}{3}{\sqrt 2} & 0 & -\frac{1}{6} 
\end{array} \right) ,
$$ 
$$
C^{0}_{1} = \left( \begin{array}{r}
\frac{1}{6}{\sqrt 2} \\ \frac{1}{4}{\sqrt 6} \\ \frac{1}{6}{\sqrt 3} 
\end{array} \right) ,
$$
$$
C^{0}_{2} = ( -\frac{1}{6}{\sqrt 2}\ \ \  0 \ \ -1/{\sqrt 3} ) ,
$$
$$
C^{1}_{1} = \left( \begin{array}{rr}
-\frac{1}{12}{\sqrt 3} & \frac{1}{4}{\sqrt 3} \\
\frac{1}{4}{\sqrt 3} & -\frac{1}{12}{\sqrt 3} \\
0 & -\frac{1}{6}{\sqrt 6}
\end{array} \right) ,
$$ 
$$ 
C^{1}_{2} = \left( \begin{array}{rrrr}
0 & -\frac{1}{3}{\sqrt 3} & 0 & \frac{1}{3}{\sqrt 6} \\ 
-\frac{1}{3}{\sqrt 3} & 0 & \frac{1}{3}{\sqrt 6} & 0 
\end{array} \right) .
$$

The expressions for the coordinates ${x_i},{y_i},{z_i}$ can be given 
by formulas  
$$
{x_1} = \left( \frac{1}{4}{x^2} + y^{2} - xyv\right) ^{1/2}, \ \ \ \ {y_1} =
\left( \frac{9}{16}{x^2} + \frac{1}{4}{y^2} + \frac{3}{4}xyv\right) ^{1/2}, 
$$
$$
{x_2} = y, \ \ {y_2} = x, \ \ {x_3} = {x_1}, \ \ {x_4} = {x_1}, \ \ {x_5} = y, 
$$
$$
{y_3} = \left(  
\frac{1}{9}{y_1}^{2}+z^{2}+\frac{2}{3}{y_1}zu\right) ^{1/2}, \ \
{z_3} = \frac{1}{3}\left(   
\frac{64}{9}{y_1}{^2}+{z^2}-\frac{16}{3}{y_1}zu\right) ^{1/2}, 
$$ 
$$
{y_4} = \left( 
\frac{4}{9}{y_1}^{2}+z^{2}+\frac{4}{3}{y_1}zu\right) ^{1/2}, \ \
{z_4} = \left( 
\frac{4}{9}{y_1}^{2}+\frac{1}{4}{z^2}-\frac{2}{3}{y_1}zu\right) ^{1/2},  
$$
$$
{y_5} = \left( \frac{1}{4}{x}^{2}+z^{2}-xzv\right) ^{1/2}, \ \
{z_5} = \frac{2}{3}\left( x^{2}+{z^2}+2xzv\right) ^{1/2}.
$$

\newpage

\vskip4cm  

\begin{tabular}{|c|c|c|c|} 
\hline 
\parbox{2cm}{\center $N$}& \parbox{3cm}{\center S3}& 
 \parbox{3cm}{\center MT I-III${^1}$} &                                
\parbox{3cm}{\center MT I-III${^2}$} \\
    &        &    &     \\ 
\hline
1    & 23.86 &  24.6945      & 25.154   \\ 
\hline 
2    & 27.11 &  28.3060      & 28.869    \\ 
\hline 
3    & 27.23 &  28.4734      & 28.975    \\
\hline 
4    & 27.75 &  28.7086      & 29.223   \\
\hline 
5    & 28.38 &  29.3345      & 29.688   \\ 
\hline 
6    & 28.42 &  29.4204      & 29.783    \\ 
\hline 
7    & 28.39 &  29.9648      & 30.284     \\ 
\hline 
8    & 28.55 &  29.5980      & 30.277     \\ 
\hline 
9    & 28.67 &  30.0534      & 30.304      \\ 
\hline 
10   & 28.72 &  30.0829      & 30.315       \\ 
\hline 
11   & 28.71 &  30.0901      & 30.307       \\ 
\hline 
12   & 28.74 &  30.1369      &         \\ 
\hline 
13   & 28.73 &  30.1392      &         \\  
\hline
\end{tabular} 

\vskip2cm

{\bf Table 1}. The convergence of the ${^4}$He ground-state energy $E$(MeV)
with respect to the number $N={N_1}={N_2}$ of terms in (\ref{symexp}) 
taken into 
account in the numerical solution of Eqs. (\ref{swaveyak}).
Here MT I-III${^1}$ is the potential from \ci{maltjo} and MT I-III${^2}$ 
from \ci{schelkok4,glock}.
\newpage 

\vskip4cm

\begin{tabular}{|c|c|c|c|}
\hline 
\parbox{3cm}{\center Refs.} & \parbox{2.5cm}{\center S3} & 
\parbox{2.5cm}{\center MT I-III${^1}$} & \parbox{2.5cm}{\center MT I-III
${^2}$}\\
               &         &          &    \\
\hline 
\ci{schelkok4} & 28.7843 &        &  30.3117      \\ 
\hline 
\ci{glock}     & 28.80\ \  &      & 30.29\ \      \\
\hline 
\ci{tjon}      &         &  29.6(2) &  \\  
\hline
present work   & 28.7(4) & 30.1(3)  & 30.3(1)    \\ 
\hline 
\end{tabular}

\vskip2cm

{\bf Table 2}. The results of the calculations of the ${^4}$He 
ground-state  
energy $E$(MeV) for spin-dependent S3 and MT I-III $N-N$ forces.
Here MT I-III${^1}$ is the potential from \ci{maltjo} and MT I-III${^2}$ 
from \ci{schelkok4,glock}.

\newpage 

\vskip4cm 
 
\begin{tabular}{|c|c|c|}
\hline
\parbox{4cm}{\center Refs.} & \parbox{2cm}{\center $A_{0}$ fm} & 
\parbox{2cm}{\center $A_{1}$ fm} \\
             &           &          \\ 
\hline 
Present work & 4.0\ \ \  & 3.6\ \ \  \\ 
\hline 
\cite{tjon} & 4.09\ \  & 3.61\ \   \\
\hline 
\cite{levashov} & 4.23\ \  & 3.46\ \   \\
\hline 
\cite{fonseca}  & 3.905 & 3.597  \\
\hline 
\cite{belyaev} & 3.8\ \  \ & 4.9\ \  \  \\ 
\hline 
\cite{heiss} & 3.38\ \  & 3.25\ \  \\
\hline 
\cite{carbon} & 4.13\ \ & 3.73\ \  \\ 
\hline 
exp. \cite{exp} & 3.91 $\pm $ 0.12 & 3.6 $\pm $ 0.1 \\   
\hline 
\end{tabular}

\vskip2cm
{\bf Table 3}. The calculated spin-singlet $A_0$ and spin-triplet $A_1$ 
scattering lengths using the MT I-III${^1}$ potential of \ci{maltjo}
, the results of other authors and experimental data 
for $n$-${^3}$H scattering.

\end{document}